\documentclass[10pt,conference,compsocconf]{IEEEtran}
\usepackage{times}
\usepackage{array}
\usepackage{amsmath}
\usepackage{amssymb}
\usepackage{Definitions}
\usepackage{caption}
\usepackage{dblfloatfix}
\usepackage{color}
\usepackage{cite}
\usepackage{algorithmic}
\usepackage{subcaption}
\usepackage{graphicx}

\begin{document}
\pagenumbering{gobble}

\title{\textbf{\Large Locating the Source in Real-world Diffusion Network}}

\author{
	\IEEEauthorblockN{Shabnam Behzad\IEEEauthorrefmark{1}, Arman Sepehr\IEEEauthorrefmark{1}, Hamid Beigy\IEEEauthorrefmark{1}, Mohammadzaman Zamani\IEEEauthorrefmark{2}} \\

	\IEEEauthorblockA{\IEEEauthorrefmark{1} Computer Engineering Department\\
		Sharif University of Technology, Tehran, Iran\\
		\{shbehzad, sepehr, beigy\}@ce.sharif.edu}  \\

	\IEEEauthorblockA{\IEEEauthorrefmark{2} Computer Science Department\\
		Stony Brook University, NY, USA
		\\mzamani@cs.stonybrook.edu}
}

\maketitle

\begin{abstract}
\boldmath
The problem of identifying the source of a propagation based on limited observations has been studied significantly in recent years, as it can help reducing the damage caused by unwanted infections. In this paper we present an efficient approach to find the node that originally introduced a piece of information into the network, and infer the time when it is initiated. Labeling infected nodes detected in limited observation as \textit{observed} nodes and other ones as \textit{hidden} nodes, we first estimate the shortest path between hidden nodes to observed ones for each propagation trace. Then we find the best node as the source among the hidden nodes by optimizing over square loss function. The method presented in this paper is based on more realistic situations and is easy and more practical than previous works. Our experiments on real-world propagation through networks show the superiority of our approach in detecting true source, boosting the top ten accuracy from less than $10\%$ for the sate-of-the-art methods to approximately $30\%$. Additionally, we observe that our source identification method runs about 10 times faster than previous work.

\end{abstract}


\IEEEpeerreviewmaketitle

\section{Introduction}

Today we are part of various social networks, and  propagation can happen from node to node rapidly over these networks. For example, information and trends may spread in social networks~\cite{bakshy2011everyone}, computer viruses may disseminate through Internet~\cite{wang2014modeling}, or even some disease may lead to an epidemic in a region~\cite{goh2006epidemiology}. The fact that some of these propagation are unpleasant and may cause enormous damages or even hurt innocent people, has urged many researchers to work on understanding the networks and propagation processes. 

One of the fundamental questions in this area is whether it is possible to find the source of these propagation, and the time when a source started to infect others. The solution for this problem has a wide range of important applications: We can predict or even prevent some rumor to spread in a social network, prevent future virus disruption, cyber-attacks, and better understand the cause or diagnose of a disease. Unfortunately, in practice, it is usually costly or even impossible to observe all nodes and trace the flow of information or virus. Hence, in such situations, we always have to deal with incomplete data, for instance we do not have full access to all the infected nodes and their infection time, or we do not know who infected whom. Thus, there should be strategies to find the source based on partially observed traces and that is what we are trying to do in this paper. 

This problem has been studied in\cite{farajtabar2015back}, in which their proposed solution worked decently on synthetic data, however its accuracy on real-world data was not pleasant. This might suggest that they were over fitted to their modeling of real-world phenomenons and consequently, misinterpreted some real-world data. In contrast our approach is more focused on being practical as well as easily implementable, even though we have experimented in settings which are much harder and more realistic than previous works~\cite{farajtabar2015back,pinto2012locating}. We have considered the case where our partial observation is only based on the final nodes of the propagation and not randomly, and with this information we try to get back to the source node.
%

The remainder of the paper is organized as follows: First ,in section~\ref{sec:related}, we revisit the related works around source identification problem. Afterwards, in section~\ref{sec:model}, we review the continuous-time independent cascade model. In sections~\ref{sec:problem}, we introduce the source identification problem, and then develop an efficient method to solve it. Then in sections~\ref{sec:experiment}, we evaluate the proposed methods using real world data. Finally, we present the conclusions
and future work in section~\ref{sec:conc}.

\section{Related Work}\label{sec:related}

Over recent years, much work has been done with different perspectives on diffusion networks. There are some works on uncovering the hidden network based on propagation traces~\cite{rodriguez2011uncovering,du2012learning,du2013uncover,
rodriguez2014uncovering,sefer2016diffusion}, the spread of influence through a network~\cite{kempe2003maximizing,du2013scalable}, reconstructing the propagation of an activity in a network~\cite{rozenshtein2016reconstructing}, and finding the source of a propagation trace~\cite{shah2011rumors,pinto2012locating,shi2017source}. 

The latter is the focus of this paper, as there are many other works in this area with a variety of noble approaches.
Some of these works are suitable for multiple source identification with different approaches: Searching for a seed set that minimizes
the symmetric difference between the cascade from seed set, and the infected nodes~\cite{nguyen2016multiple}, or extracting a sub-graph using candidate selection algorithm and then presenting OJC algorithm, which finds a set of nodes that cover all observed infected nodes with the minimum radius~\cite{zhu2017catch}.
Employment of the minimum description length principle is another approach used by \cite{prakash2012spotting}. Moreover, researchers proposed methods that first injects sensors into
networks and then identifies the propagation sources~\cite{aldalahmeh2011robust,bianchi2011performance}.

The work most closely related to ours, however, is \cite{farajtabar2015back} which first infers networks through cascade
data and then tries to find the best source by maximizing the likelihood of traces under the learned model. Unfortunately, we find the performance of presented algorithm underwhelming, and in this paper, we tried to boost the accuracy of source identification with a new approach. Additionally, in \cite{farajtabar2015back}, it takes a vast amount of time to converge to the solution value \eg more than a day for a graph with 1024 nodes, while we are trying to find the source in a reasonable amount of time, which makes our approach applicable to even larger networks that we already have data for.

\section{Continuous Time Diffusion Model}\label{sec:model}

Let $G=(\nu,\xi)$ be a directed graph. When a node, $s \in \nu $, begins a rumor or starts spreading a virus at time $t_{s}$, a propagation process begins. Nodes can transmit the contagion along their out-going edges to their directed neighbors, in other words, node $j$ can infect node $i$, if there is an edge from $j$ to $i$. A node can have one of the two possible states: i) \textit{infected}, if it has received the information or virus within a certain time window; or ii) \textit{ignorant}, if it has not been infected in a specific time window\cite{pinto2012locating}. Each infected node may infect its neighbors with a random spreading time $\tau_{ij}=t_i-t_j$ in the case that node $j$ is true parent of node $i$. The spreading time is drawn from a density over time for each edge,  represented in terms of a parameter, $\alpha_{ij}$, which stands for the transmission rate of each edge. These transmission rates are independent, and we assume that a node can not recover from infection. Also, an infected node can not get infected more than one time. 

Given an observation window of length $T$, a propagation or cascade $C_{c}= <t_{c}^{1}, \dots , t_{c}^{\mathcal{V}}>$
is a $\mathcal{V}$-dimensional vector recording the time when each of the nodes in the network got infected in the specific time window. In other words, $t_{c}^{i}$ indicates the infection time of node $i$ in cascade $c$. We can label nodes as \textit{ignorant} if infection time is $\infty$, which means they were not infected during the observation time. In this paper, we assume that we do not have all of the infected nodes. Thus, we have two sets of nodes, the \textit{observed} ones, $\mathcal{O}$, with their infection time, and the \textit{hidden} nodes, $\mathcal{H}$. Since we assume that our data collection starts at some time after the starting point of the propagation, we can be sure that $s\in \mathcal{H}$, so we try to find the source among hidden nodes. To this end, we found it helpful to find the source node, and its infection time, rather than just looking for the most probable source node.

\section{Problem Formulation}\label{sec:problem}

Consider the case when there is a set of cascades from the same source at different times, 
but we only know about some of the infected nodes and their infection time in each cascade. We call these nodes the observed ones, and we collect them in two different settings. In the first one, we consider the case when the infection has spread for a while, and after some time we listen on the whole network and observe all the nodes that has been infected since then and record their infection times. In the second setting, we observe some nodes randomly from the start of the propagation and record their infection times. In both scenarios, we have two goals at the same time, the first one is to find the starting time of each propagation, knowing the fact that for different propagation traces from a specific source, we may have different starting times and we want to find them all separately. The second goal which is the main one, is that for each set of cascades, we would like to find the joint node who initiated all those cascades.

\subsection{Our Approach}\label{sec:approach}

\begin{figure*}[t]\label{fig1}
\centering
\begin{subfigure}{.45\textwidth}
  \centering
  \includegraphics[width=0.9\linewidth]{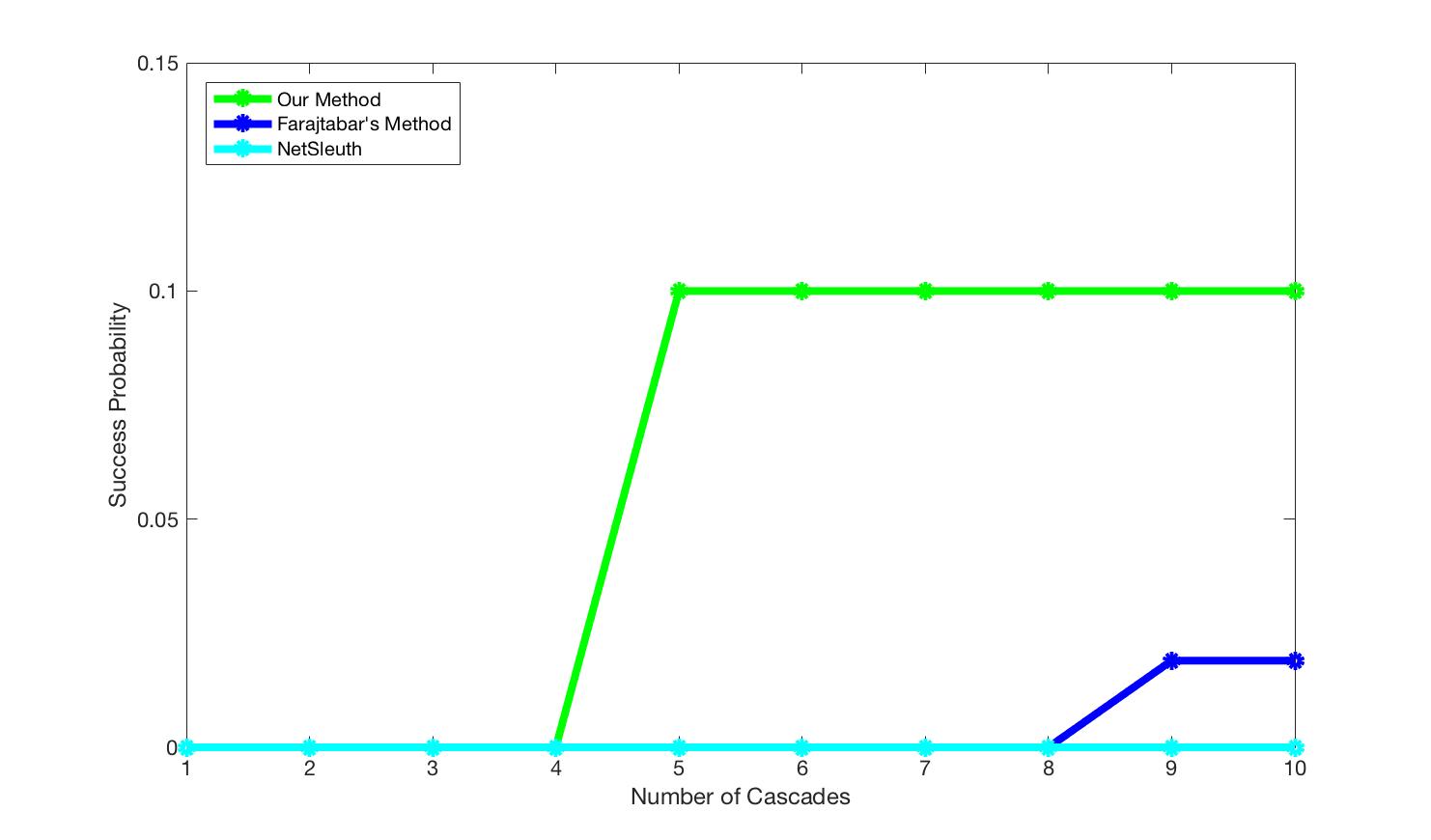}
  \caption{Success Probability}
  \label{fig:sub1}
\end{subfigure}%
\begin{subfigure}{.45\textwidth}
  \centering
  \includegraphics[width=0.9\linewidth]{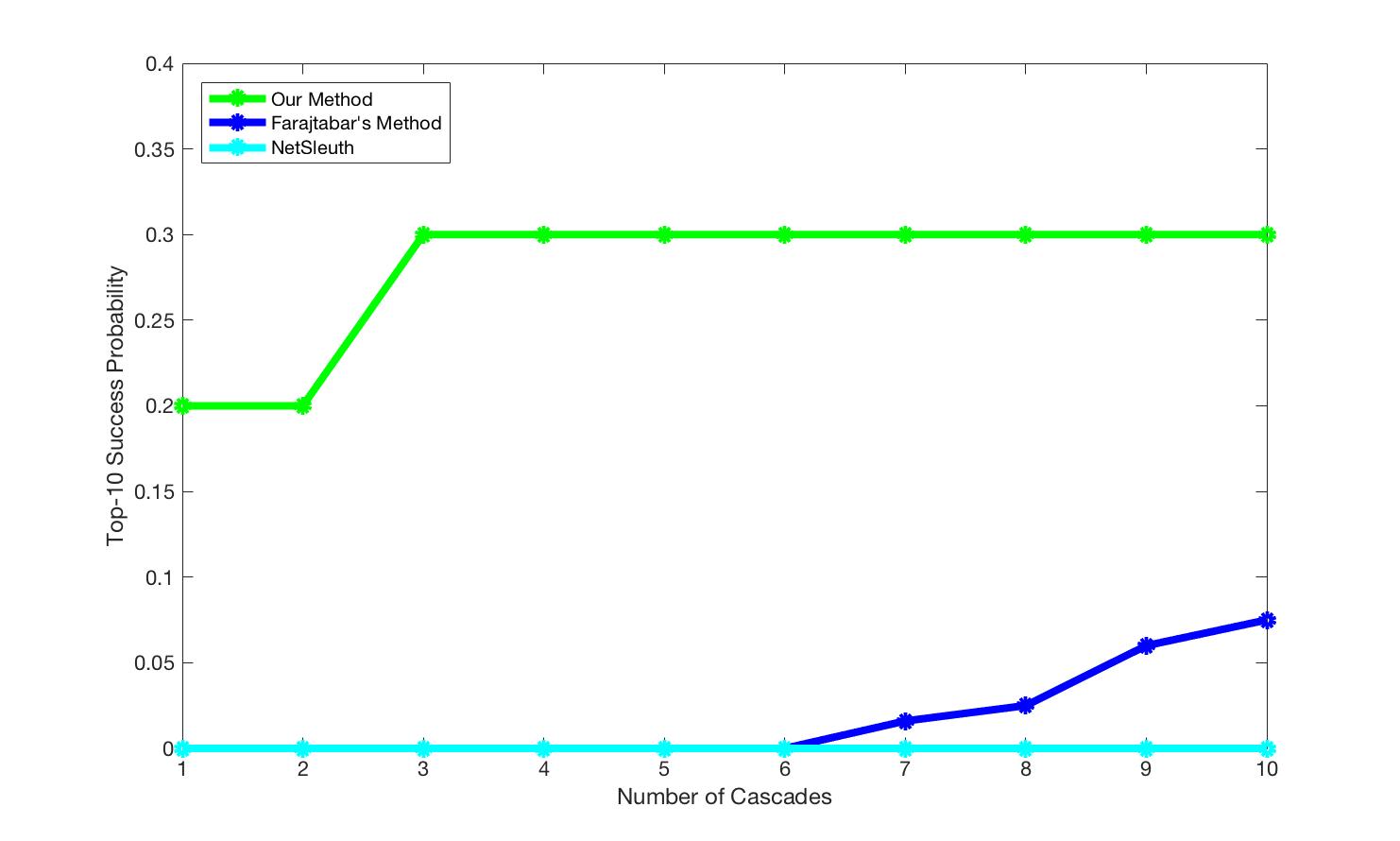}
  \caption{Top-10 Success Probability}
  \label{fig:sub2}
\end{subfigure}
\caption{Success Probability and Top-10 Success Probability for real propagation data with random observed nodes.}
\label{fig:test}
\end{figure*}

Our approach consists of two main steps: First we try to reconstruct the network by a fair number of cascades, and then we try to find the source, $s$, and the infection time of the source $t_{s}$.
For each partially observed cascade $i$, we have a set $\mathcal{O}_{c}$ which indicates the observed nodes of that cascade. This means that we have the infection time of all the nodes in $\mathcal{O}_{c}$. The source, however, is in the hidden nodes $\mathcal{H}_{c}$, and $|\mathcal{O}_{c}|=|\mathcal{V}|-|\mathcal{H}_{c}|=K$. 

Our approach builds up on three parts which are introduced in detail: We will reconstruct the network based on cascades, then we will estimate the shortest path with a novel approach. Finally, with estimated values, we minimize an objective function to find the best source node.

\subsubsection{Reconstructing the Network}

First, having a fair number of historical cascades of the network, we try to reconstruct the network and find the transmission rates. We do this by implementing the \textsc{Net\-Rate} algorithm. In \cite{rodriguez2011uncovering}, it is shown that, the the likelihood of the infections in a cascade is:
\begin{multline}\label{0}
f(t_{i}|\{t_{j}\}_{j \in \pi_{i}})=\prod_{t_i\leq T} \prod_{t_m>T}S(T|t_i;
\alpha_{i,m})\\
\prod_{k:t_k<t_i}S(t_i|t_k;\alpha_{k,i})\sum_{j:t_j<t_i}\frac{f(t_i|t_j;\alpha_{j,i})}{S(t_i|t_j;\alpha_{j,i})}
\end{multline}

Where $S(t_i|t_j;\alpha_{j,i})=1-F(t_i|t_j;\alpha_{j,i})$ and $F(t_i|t_j;\alpha_{j,i})$ is the cumulative density function which is computed from the transmission likelihoods. But in our case, we assume that we have partially observed nodes and only a subset $\mathcal{O}$ of the infected nodes are observed. \cite{farajtabar2015back} showed that the likelihood of the incomplete cascade is computed as follows,
\begin{multline}\label{1}
p(\{t_i\}_{i \in \mathcal{O}}|t_s)= \int_{\Omega} p(t|t_s) \prod_{j \in \mathcal{H}} dt_j\\
=\int_{\Omega} \prod_{i \in \mathcal{O} \cup \mathcal{H}} p(t_i|\{t_j\}_{j \in \pi_i})
\prod_{j \in \mathcal{H}} dt_j
\end{multline}

Furthermore, we worked with the well-known exponential distribution which is included in the Weibull family of distributions $f_{ji}=(\tau,\alpha_{ji})$, and has been showed to fit well in real world propagation data \cite{du2013uncover,farajtabar2015back}.

\subsubsection{Estimating the Shortest Path}

After reconstructing the network, we try to find the shortest path between all hidden nodes and observed ones. Let $\Qcal_i(\Acal)$ be the collection of directed paths in $\Gcal$ from source nodes $\Acal$ to node
$i$, where each path $q\in \Qcal_i$ contains a sequence of directed edges $(j,l)$. Assuming all source nodes are infected at time zero, then we obtain variable $\hat{t_i}$ -estimated infection time for node $i$- via
\begin{equation} \label{eq:mapping}
  \hat{t_i} = g_i\rbr{\{\tau_{ji}\}_{(j,i)\in \Ecal} | \Acal} := \min_{q \in \Qcal_i(\Acal)} \sum\nolimits_{(j,l)\in q} \tau_{jl},
\end{equation}
where the transformation $g_i(\cdot)$ is the value of the shortest-path minimization.

We find the transmission time $\tau_{jl}$ for each edge by importance sampling. Due to the fact that we can not evaluate analytically the integral in Eq.~(\ref{1}), we use Monte Carlo approximation because it will converge to the true value with sufficient number of samples.
Finally, we can estimate the shortest path between all observed and hidden nodes for each sample using Dijkstra algorithm.

As we wanted our method to result in a reasonable amount of time, we have reversed the edges in the network, and we calculated the shortest path from observed nodes to hidden ones. This improves the average run time significantly, since the number of observed nodes is much smaller than the number of hidden ones. 

\subsubsection{Objective Function}

Assuming we have a set of cascades with the same source, we try to find the source for this set. This does not imply that the starting time for each cascade is the same as others. Furthermore, in this paper, we supposed that for each cascade, only one node is responsible for initiating the propagation process.

For each cascade $c$, we have a set of observed nodes $\mathcal{O}_{c}$. As described before, we have calculated the shortest path between all observed nodes and hidden ones, and that is our estimated infection time $\hat{t_i}$ of the observed node $i$ with some candidate node $s\in \mathcal{H}$ as the source. One should take into consideration the fact that in our sampled network, $\hat{t}_{s}=0$. Thus, there should be a difference between real infection time of an observed node and the estimated infection time. This difference is approximately the infection time of the source, $t_{s}$. Hence, for each candidate source $s$ in cascade $c$, we would have:

\begin{equation}\label{2}
\underset{t}{\operatorname{argmin}} \sum_{i\in \mathcal{O}_{c} }(t_{i}-\hat{t_{i}}-t_{s})^2
\end{equation}

With $t_{i}$ being the real infection time of observed node $i$ and $\hat{t_{i}}$ the estimated infection time for node $i$. Thus, with taking the derivative of Eq.~(\ref{2}):

\begin{equation}
\sum_{i\in \mathcal{O}_{c} } t_{i}-\sum_{i\in \mathcal{O}_{c} }\hat{t_{i}}-Kt_{s}=0
\end{equation}

Note that $K=|\mathcal{O}_{c}|=|\mathcal{V}|-|\mathcal{H}_{c}|$.
Now for each cascade, we can find $t_{s}$ for each candidate source via:

\begin{equation}
t_{s}=\frac{\sum_{i\in \mathcal{O}_{c} } t_{i}}{K} - \frac{\sum_{i\in \mathcal{O}_{c} } \hat{t_{i}}}{K}
\end{equation}

Which means we need to calculate the average of the real infection times and, the average of estimated infection times among observed nodes in each cascade to obtain $t_s$.

Now that we have $t_{s}$ for each candidate source in each cascade, finally we can choose the best candidate source for a set of cascades $C$, by minimizing the square error for each of the set's cascades:

\begin{equation}\label{obj}
\underset{\mathcal{H}_{C}} {\operatorname{argmin}} \sum_{c \in C}  {(\sum_{i \in \mathcal{O}{c}} (\hat{x_{i}^{c}} + t_{s}^{c}-x_{i}^{c}) )}^2
\end{equation}

Where $x_{i}^{c}$ and  $\hat{x_{i}^{c}}$ are average of the real infection times, and estimated infection times correspondingly, $\mathcal{H}_{C}=\mathcal{V}-(\mathcal{O}_{1} \bigcup \dots \mathcal{O}_{m-1}\bigcup \mathcal{O}_{m})$ and, $m$ is the number of cascades in the set.

\section{Experiments and Results}\label{sec:experiment}

Our experiments in this paper are on real data sets. Our focus is on the flow of memes which can act as signatures of topics, events, and diffuse over the web~\cite{leskovec2009meme}. The data set consists of cascade data for different topics and world news for the 5,000 most active sites from four million sites \cite{gomez2013structure , data}. In this large data set, there are a great number of cascades per topic and each cascade consist of a specific meme's propagation trace, \ie the time when different blogs has mentioned that meme.

We proceed as follows: We first estimate the underlying network using \textsc{Net\-Rate}~\cite{rodriguez2011uncovering} with all the historical cascades. Then based on this network, we try to estimate the source of a cascade. We do this by choosing a number of sources which have at least 10 long cascades, \ie consisting of more than 27 nodes. We employed 500 samples and we have two different approaches for selecting the observed nodes.

In the first set of experiments, we choose 10 different sources. The selection of the observed nodes is randomly in this setup as in \cite{farajtabar2015back}. We assume that for each cascade, only $10\%$ of the nodes are observed. As mentioned before, we have selected 10 sources with at least 10 long cascades. Hence, for each source we have different sets of cascades. These sets comprise of a minimum and maximum of 1 and 6 cascades, respectively. We evaluate the results of our approach based on two different criteria: Success Probability, which is the probability of finding the correct source and, Top-10 Success probability, which is the probability of the real source ranked among the top 10 estimated sources based on Eq.~(\ref{obj}). We have compared our results with Farajtabar's method \cite{farajtabar2015back} and \textsc{Net\-Sleuth} \cite{prakash2012spotting}. Surprisingly, their methods can not detect the source effectively. However, we have a $10\%$ success probability having only 5 cascades and a $20\%$ top-10 success probability with only 1 cascade. The results are summarized in Figure 1. It is also important to mention that while our top-10 success probability is the same for 3 to 10 cascades, we are getting more accurate ranking results for the real source. As an example, given 8 cascades for each source, the \textit{top-5} success probability is $30\%$, however, we are working on a better criterion than SE in Eq.~(\ref{obj}) to enhance the performance, when the number of the given cascades is increasing.

In the second approach, we try to simulate a harder but more realistic situation. Consider the following scenario: When there is a report of an infection at time $T$, we start to observe the propagation. Hence, we do not have any information about the nodes who got infected before $T$, we have only observed the final nodes who got infected and we want to trace them back to the source. In this case, for each cascade, $\mathcal{O}{c}$ consists of the final nodes of each cascade. Again, we consider that $10\%$ of the cascade nodes are observed and that each set comprise of a minimum and maximum of 8 and 10 cascades, respectively. We selected 5 sources for this setup and our approach is working effectively even in this scenario and, it has a top 10 success probability of approximately $35\%$.

We still have other experiments which are in progress and we are going to evaluate more results. We would like to detect how our approach works if we consider more than $10\%$ of the nodes,observed. Moreover we are going to run our algorithm on some synthetic data and assess the results.

\section{Conclusion}\label{sec:conc}
In this paper, we referred to the challenging problem of finding the source of a propagation -cascade- in a diffusion model. Our approach is based on partially observed cascades. We have two cases for the observed nodes: Random observed nodes, or the final nodes of the propagation as observed nodes. We first infer the underlying network, then we propose to employ a shortest path algorithm and find the estimated infection time of the observed nodes for each candidate source. Finally, we minimized the squared error to find the best source among all nodes. To the best of our knowledge, our approach is performing better considering accuracy,simplicity and timing, compared to the previous state-of-the-arts.

Future work can include extending our method to the SIR model or, focusing on approaches that can identify multiple sources of propagation. Additionally, there is the interesting problem of finding the sufficient number of cascades and observed nodes which will result in finding the source node, correctly. 

\
\bibliographystyle{IEEEtran}
\bibliography{refs}

\end{document}